\documentclass[pageno]{jpaper}

\usepackage[normalem]{ulem}
\usepackage{multirow}
\usepackage{xspace}
\usepackage{fancyvrb}
\usepackage{amsfonts, amsmath, amsthm, amssymb}

\usepackage{soul}

\usepackage{physics}
\usepackage{syntax}
\usepackage{listings}
\usepackage{xcolor}
\usepackage{geometry}
\usepackage{mathpartir}
\usepackage[subtle]{savetrees}
\usepackage{algorithm}
\usepackage{graphicx}
\usepackage[noend]{algpseudocode}
\usepackage{circuitikz}
\usepackage{enumitem}
\usepackage{tikz}
\usetikzlibrary{arrows,shapes,snakes,automata,backgrounds,petri,graphs,quotes}
\usepackage{wrapfig}
\usepackage{mathtools}
\usepackage{amsmath}
\usepackage{amssymb}
\usepackage{array,etoolbox}
\usepackage{multirow}
\newcommand{\thoughts}[1]{{\color{forestgreen}Thoughts: #1}}

\newcommand{\tightbullet}{\noindent}
\newcommand{\proseheading}[1]{\tightbullet{\textbf{{#1}}}}

\definecolor{deepred}{HTML}{EA2027}
\definecolor{deepblue}{HTML}{1B1464}
\definecolor{forestgreen}{HTML}{006266}
\definecolor{ultraviolet}{HTML}{341f97}
\definecolor{seafoam}{HTML}{1289A7}
\definecolor{radiantyellow}{HTML}{F79F1F}
\definecolor{burntorange}{HTML}{d35400}
\definecolor{darkgrey}{HTML}{273c75}
\definecolor{pastelblue}{HTML}{E3F4F4}

\newcommand{\dg}{\textsc{DG}\xspace{}}
\newcommand{\tool}{\textsf{Ark}\xspace{}}
\newcommand{\model}{DG\xspace{}}
\newcommand{\computingparadigm}[1]{#1}
\newcommand{\gmctln}{\computingparadigm{GmC-TLN}}
\newcommand{\tln}{\computingparadigm{TLN}}
\newcommand{\cnn}{\computingparadigm{CNN}}
\newcommand{\con}{\computingparadigm{OBC}}
\newcommand{\crn}{\computingparadigm{CRN}}

\newcommand{\puf}{PUF}

\newcommand{\syn}[1]{\texttt{\textcolor{deepred}{#1}}}

\newcommand{\lines}[1]{\textcolor{burntorange}{#1}}

\newcommand{\nonnegativenat}{\mathbb{N}_0}
\newcommand{\realnumbers}{\mathbb{R}}
\newcommand{\positivereal}{\mathbb{R}^{++}}
\newcommand{\integers}{\mathbb{Z}}

\newcommand{\tuple}[1]{\langle #1 \rangle}
\newcommand{\set}[1]{\{ #1 \}}
\newcommand{\cmt}[1]{\textcolor{gray}{#1}}

\newcommand{\litl}[1]{\texttt{\textcolor{forestgreen}{#1}}}

\newcommand{\codein}[1]{{\textbf{\texttt{#1}}}}

\newcommand{\nodetype}{NT}
\newcommand{\edgetype}{ET}
\newcommand{\nodeorder}{p}

\newcommand{\othernode}{\Hat{\node}}
\newcommand{\node}{n}
\newcommand{\nodeset}{\mathtt{N}}

\newcommand{\edge}{e}
\newcommand{\edgeset}{\mathtt{E}}

\newcommand{\hwst}[1]{\colorbox{pastelblue}{#1}}

\makeatletter
\DeclareRobustCommand\bigop[1]{%
  \mathop{\vphantom{\sum}\mathpalette\bigop@{#1}}\slimits@
}
\newcommand{\bigop@}[2]{%
  \vcenter{%
    \sbox\z@{$#1\sum$}%
    \hbox{\resizebox{\ifx#1\displaystyle1\fi\dimexpr\ht\z@+\dp\z@}{!}{$\m@th#2$}}%
  }%
}
\makeatother

\newcommand{\bigLambda}{\DOTSB\bigop{\Lambda}}

\newcommand{\reduction}{\bigLambda}
\newcommand{\ddt}[1]{\frac{d#1}{dt}}
\newcommand{\cint}{C_{int}}
\newcommand{\gint}{G_{int}}
\newcommand{\gm}{Gm}
\newcommand{\norder}[1]{#1\textsuperscript{th}}
\tikzset{scaledgm/.style={muxdemux, muxdemux def={NL=2, Lh=3, NR=1, Rh=1, NB=0, w=1}, font=\scriptsize}}

\newcommand{\cnnexp}[1]{{\large\texttt{{#1}}}}

\newcommand{\maxcut}{max-cut}

\newcommand{\algkw}[1]{\textbf{#1}\xspace{}}
\newcommand{\algfnprose}[1]{\texttt{#1}}

\newcommand{\algmatchprod}{\algkw{LookUpProdRule}}
\newcommand{\algapp}{append}
\newcommand{\algnodes}{\algkw{NodesOf}}
\newcommand{\algedges}{\algkw{EdgesOf}}

\title{Design of Novel Analog Compute Paradigms with \tool{}}
\author{Yu-Neng Wang$^1$, Glenn Cowan$^2$, Ulrich Rührmair$^{3,4}$, Sara Achour$^1$\\
$^1$Stanford University, $^2$Concordia University, $^3$TU Berlin, $^4$University of Connecticut\\
\texttt{\{wynwyn,sachour\}@stanford.edu, gcowan@ece.concordia.ca, ruehrmair@ilo.de}}
\date{}

\preto\tabular{\setcounter{magicrownumbers}{0}}
\newcounter{magicrownumbers}
\newcommand\rownumber{\stepcounter{magicrownumbers}{\arabic{magicrownumbers}}}

\begin{document}

\maketitle

\begin{abstract}

{Previous efforts on reconfigurable analog circuits mostly focused on specialized analog
circuits, produced through careful co-design, or on highly
reconfigurable, but relatively resource inefficient, accelerators
that implement analog compute paradigms. This work deals with
an intermediate point in the design space: Specialized reconfigurable
circuits for analog compute paradigms. This class of
circuits requires new methodologies for performing co-design,
as prior techniques are typically highly specialized to conventional circuit classes (e.g., filters, ADCs).

In this context, we present \tool{}, a programming language for describing
analog compute paradigms. \tool{} enables progressive
incorporation of analog behaviors into computations, and
deploys a validator and dynamical system compiler for
verifying and simulating computations. We use \tool{} to codify
the design space for three different exemplary circuit design problems,
and demonstrate that \tool{} helps exploring design trade-offs and
evaluating the impact of nonidealities to the computation.
 }


\end{abstract}
\section{Introduction}

There has been significant interest in domain-specific reconfigurable analog computing platforms that perform in-sensor and near- or in-memory computation and solve computationally hard problems~\cite{ryynanen2001dual,gangopadhyay2014compressed,mehonic2020memristors,konatham2020real,sebastian2020memory}. These analog architectures have applications in machine vision, medical devices, and robotics domains,~\cite{jang2020atomically, northrop2003analysis, koziol2012robot} and reduce data movement across the analog/digital interface, enabling the use of more resource-efficient digital hardware~\cite{Decadal, murmann2020a2i}. These analog architectures are implemented in circuits with good performance and typically incorporated into a larger domain-specific computation: examples include scientific computation and sensor processing pipelines.~\cite{cowan2005vlsi, huang2017hybrid, tsividis2018analog-computer, guo2016hybrid-computer}

\proseheading{Classical Analog Circuits.} Designers have developed highly specialized classical analog circuits (e.g., filters, ADCs) with programmability and fidelity characteristics tailored for specific application use cases. These circuits typically offer limited programmability and target a specific fidelity and therefore can be engineered to be highly resource-efficient. Because these circuits are largely non-programmable, much of the specialization is done in the design stage. Fortunately, the design requirements for classical circuits can be described with standard figures of merit and functional specifications, and circuit designers have a good intuition on how to craft an implementation that meets the requirements. The existence of these standard interfaces between hardware designers and domain specialists is critical; otherwise, design-time specialization would be highly challenging.



\proseheading{Unconventional Analog Accelerators.} Researchers have developed highly reconfigurable analog accelerators that faithfully implement radical new forms of computation, such as GPAC computing, oscillator-based computing, spiking neural networks, and cellular non-linear networks.~\cite{cowan2005vlsi,huang2017hybrid,chua1988cellular-theory,ghosh2009spiking,csaba2020coupled,yakovlev2020tln-compute} Many of these accelerators require  unconventional uses of analog circuits to implement novel computational operators or compute on  different signal components (e.g., transient behavior). Therefore, these accelerators faithfully implement their respective analog compute paradigms and are highly programmable. Supporting this degree of generality comes at a substantial resource and performance cost in analog design, impacting the hardware platform's scalability and efficiency. In contrast to specialized analog circuits, these accelerators are not typically developed through careful co-design, as that hardware is sufficiently flexible to support computations that are expressible in the target compute paradigm.

\subsection{Domain-Specific Unconventional Analog Circuits} 

\begin{figure}[t!]
\includegraphics[width=\linewidth]{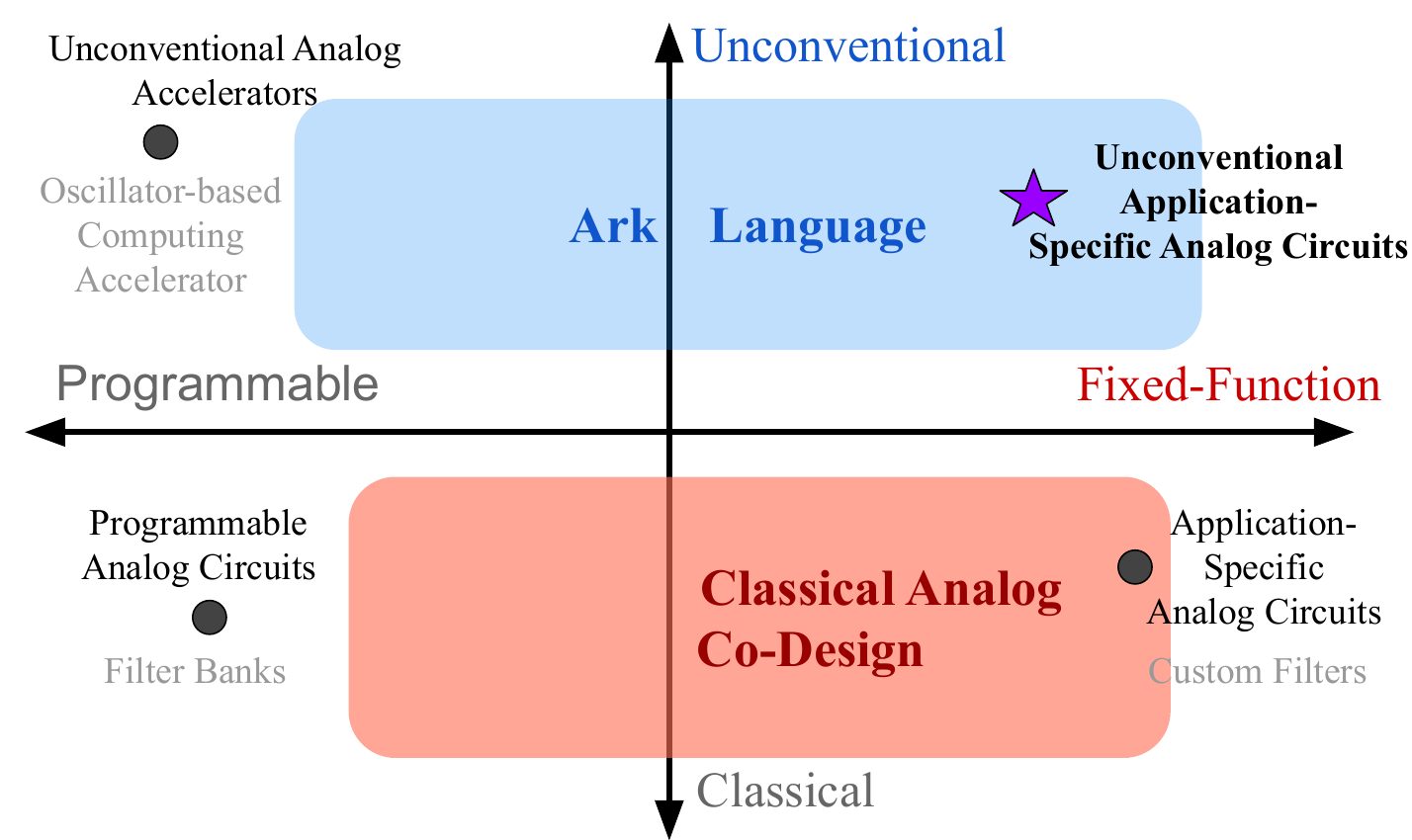}
\caption{Analog design space exploration.}
\label{fig:intro:DSE}
\end{figure}

Thus far, prior work has focused on two extremes in the analog design space (1) highly specialized classical circuits and (2) highly programmable analog accelerators that target unconventional analog compute paradigms. We anticipate there is a largely untapped part of this design space: highly specialized analog circuits that perform computation using an unconventional analog compute paradigm (Figure~\ref{fig:intro:DSE}). Because these circuits are specialized to a particular application domain, the circuit's programmability and fidelity requirements can potentially be reduced, lowering resource costs and improving performance.

\proseheading{Design Challenges.} To realize this new part of the design space, we need to specialize unconventional analog circuits to the target application domain. However, because these designs are typically non-standard, methods for specifying functional behavior and design requirements for classical circuits may not sufficiently capture the relevant behaviors and design requirements. Second, there is less design intuition for unconventional circuits, as they haven't been studied as extensively as ADCs, filters, and other classical analog circuits.  Therefore, it is highly challenging for the domain specialist and analog designer to navigate this space effectively and arrive at a promising design.


\subsection{Analog Design with \tool{}}

We introduce \tool{}, a programming language that enables design space exploration of unconventional analog circuits that leverage analog compute paradigms. With \tool{}, analog designers can expose different programmability and fidelity tradeoffs for domain specialists to explore within the context of their workload. The \tool{} language offers the following capabilities:

\begin{itemize}[leftmargin=*]
\item{}\textbf{Analog Compute Paradigms as DSLs}: \tool{} supports the specification of new analog compute paradigms as domain-specific languages (DSLs), and deploys a compiler that enables the validation and simulation of computations written in an analog compute DSL.

\item{}\textbf{Codifying Analog Design Spaces with DSLs}: With \tool{}, analog designers can specialize a compute paradigm DSL to incorporate new language constructs that capture bottom-up analog design constraints, model nonidealities, and codify different design tradeoffs. 

\item{}\textbf{Progressive Design of Analog Circuit Descriptions}: \tool{} supports the specification, configuration, and simulation of reconfigurable analog computations. \tool{} imposes strict inheritance rules to ensure computations written in the original compute paradigm  can be progressively rewritten to selectively incorporate analog behaviors.
\end{itemize}

\noindent\textbf{Design Flow.} With \tool{}, both the domain specialist and the analog designer settle on an analog compute paradigm to serve as a basis for building both analog circuit designs and computations. Each analog compute paradigm offers basic computational operators that modify certain properties (e.g., phase) of the underlying analog signal and can be simulated with a system of differential equations. The domain specialist first develops reconfigurable analog computations in the agreed upon analog compute DSL for their target workloads, and the analog designer extends the analog compute DSL to codify the analog design space. 

The domain specialist then progressively adapts computations to use designer-provided constructs to analyze the effect of analog nonidealities in their computation and explore different analog design options. The analog designer may then declare new constructs that codify new design points based on the domain specialist's usage of their language extension. Reconfigurable analog computations defined entirely with analog hardware constructs serve as a functional and requirements specification for unconventional analog circuits. We anticipate this flow enables iterative co-design of analog circuits.


\subsection{Contributions}

\begin{itemize}

\item{}\textbf{Dynamical Graph Representation}: We present a novel, unified intermediate representation termed a dynamical graph (\dg{}) for both analog computations and analog circuit descriptions.

\item{}\textbf{The \tool{} Language}: We present a programming language that supports the definition of domain-specific languages that codify analog compute paradigms and their associated hardware design spaces. Reconfigurable analog computations can then be written in DSLs in the defined language.

\item{}\textbf{\tool{} Compiler and Validator}: We present a compiler that derives the system of differential equations that simulates the transient dynamics of a given \tool{} program and a validator that verifies that a given \tool{} program satisfies all of the constraints imposed by the domain-specific language.


\item{}\textbf{Case Study and Evaluation}:  We use \tool{} to codify the design tradeoffs associated with a transmission-line based analog \puf{} (\tln{}), a cellular non-linear network (\cnn{}) analog accelerator, and an oscillator-based computing (\con{}) analog accelerator. We demonstrate that \tool{} can capture nonidealities and design tradeoffs associated with these design problems and provide a detailed analysis for the \puf{} design problem.

\end{itemize}

\section{Case Study: Transmission Line PUF}

\begin{figure*}[!tb]
    \centering
    \includegraphics[width=0.95\linewidth]{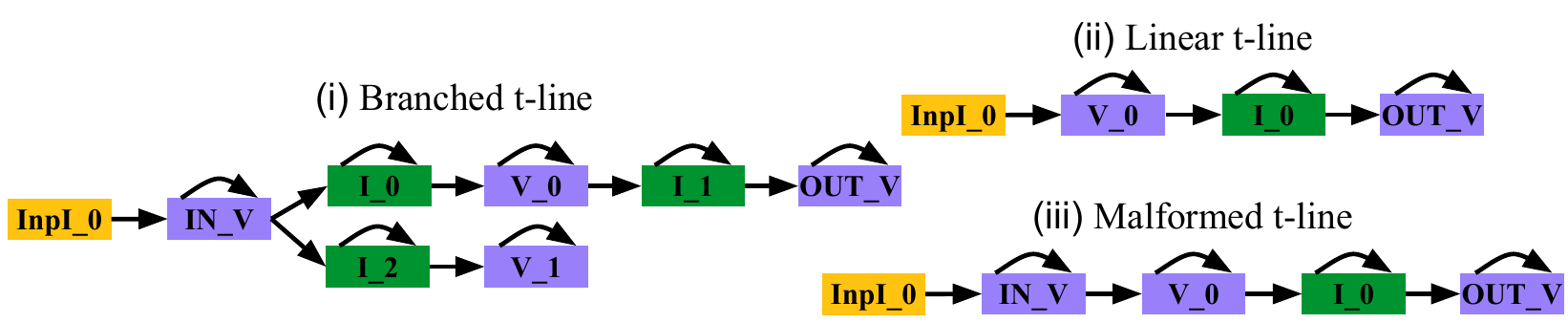}
    \caption{Dynamical graphs of branched, linear, and malformed t-lines.}
    \label{fig:example:tln:dgs} 
\end{figure*}
\begin{figure}[!tb]
    \centering
    \begin{circuitikz}[american voltages]
\ctikzset{resistors/scale=.7,
          capacitors/scale=0.7
          }
\draw
    (3,0.2) to [short] (4,0.2)
    to [R, a=$\gint$, font=\footnotesize] (4,2) 
    to [short] (3,2)
    to [C=, l=$\cint$, v=$v_o$, font=\footnotesize] (3,0.2)
    (0,1) node[scaledgm] (gm1) {$\gm_1$}
    node [left, font=\footnotesize] at (gm1.lpin 1) {+}
    node [left, font=] at (gm1.lpin 2) {-}
    node [left, font=\footnotesize] at (gm1.west) {$v_{i1}$}
    (1.5,1) node[scaledgm] (gm2) {$\gm_2$}
    node [left, font=\footnotesize] at (gm2.lpin 1) {+}
    node [left, font=] at (gm2.lpin 2) {-}
    node [left, font=\footnotesize] at (gm2.west) {$v_{i1}$}
    (gm1.rpin 1) |- (3, 2)
    (gm2.rpin 1) |- (3, 2)
    ;
\end{circuitikz}
    \caption{A GmC integrator schematic.}
    \label{fig:example:gmc-integrator} 
\end{figure}
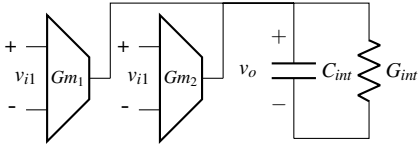


We start with an illustrative example to give an overview of the language components in \tool{} and demonstrate the design flow enabled by \tool{}.
This case study focuses on the design of an unconventional analog circuit that implements a \textit{physical unclonable function} (\puf)~\cite{herder2014puf-tutorial}, a security primitive that leverages fabrication variations to produce a difficult-to-spoof hardware authentication device. Given a challenge bitvector, a \puf{} produces a response bitvector that is highly sensitive to fabrication variation.  In a well-designed PUF, the mapping between challenge and response should be stable but maximally complex and hard to imitate or predict for cryptographic adversaries without physically possessing and interrogating the PUF.   In an analog circuit \puf{}, the response is often naturally computed from voltage and current trajectories observed on a wire within a certain observation time window.

The security expert opts to target the \textit{transmission line network (\tln{}) compute paradigm}~\cite{csaba2009cnn-puf} and selects the \tln{} DSL (Section~\ref{subsec:lang:tln-example}) provided by \tool{}. To investigate the effect of fabrication variation, the expert uses the \gmctln{} extension of the \tln{} DSL (Section~\ref{subsec:lang:gmc-tln}), which codifies the design space of fabrication-variation-sensitive GmC circuit implementations. 

\subsection{The \tln{} Compute Paradigm}\label{subsec:example:tln-computational}
A \textit{transmission line} (t-line) is a channel that carries electromagnetic waves across some distance. The traversal of a wave through a line is modeled with the discretized \textit{Telegrapher's equations}~\cite{inan2016engineering-em-wave}:
\begin{equation}
    \begin{cases}
         \frac{dV_{i}}{dt} = \frac{1}{C_i}( I_{i} - I_{i+1} - G\cdot V_{i})\\
        \frac{dI_{i}}{dt} = \frac{1}{L_i} (V_{i-1} - V_{i} - R\cdot I_{i})
    \end{cases}
    \label{eq:example:telegrapher}
\end{equation}
The t-line is segmented into 0.., $i$,...$n$ segments, where $V_i$ and $I_i$ models the voltage and current at each line segment $i$. The $R$ and $G$ parameters model impedance, which attenuates signals, and $L$ and $C$ parameters model propagation speeds, which delay signals and set the characteristic impedance. A \textit{transmission line network} is a network of interconnected t-lines that route signals and leverage the delay, reflection, and transmission for computation. \textit{Reflection} and \textit{transmission} occur when the characteristic impedance is changed within a line, such as when branches are introduced or when a line is terminated.



\subsection{Exploring \tln{} Topologies with \tool{}}\label{subsec:example:explore-topo}

\tool{} provides a \tln{} DSL that implements transmission line networks modeled with the Telegrapher's equations. Figure~\ref{fig:example:tln:dgs}-(i) presents a branched t-line network implemented in the \tln{} DSL and formulated as a dynamical graph (Section~\ref{sec:dg}). A dynamical graph is a typed, directed graph with nodes and edges that map to variables and dynamics in the underlying dynamical system, respectively. In the branched t-line, the \codein{V}, \codein{I} node types (purple, green) map to $V_i$ and $I_i$ variables in the Telegrapher's equation model, and the and \codein{InpI} node type (yellow) injects an external signal into the transmission line network.

\proseheading{Attributes.} Each node and edge type defines attributes fixed to values at simulation time. The \codein{V} and \codein{I} node types define real-valued \codein{c}/\codein{g} and  \codein{l}/\codein{r} attributes, respectively, which map to the $C$, $G$, $L$ and $R$ parameters in the Telegrapher's equations. In the above example, all \codein{l}, \codein{c} attributes are set to \codein{1e-9}, and all \codein{g},\codein{r} attributes are set to \codein{0} for nodes in the middle of the line and \codein{1} for the \codein{IN_V} and \codein{OUT_V}. The \codein{InpI} node type defines a function attribute that is assigned an input pulse function \codein{pulse(t,0,2e-8)} in the branched t-line. The node and edge types and attributes are defined in \tln{} DSL.

\proseheading{Dynamics.} In the dynamical graph, interactions are \textit{local}, and neighboring edges contribute terms to the differential equations. In branched t-line, the incoming, outgoing, and self-referencing edges contribute $I/V.c$, $-I/V.c$, $-V.g/V.c \cdot V$ terms to each $V$ node's differential equations, and contribute $V/I.l$, $-V/I.l$, $-I.r/I.l \cdot I$ terms to each $I$ node's differential equations. The \tln{} language defines the production rules for translating connections to algebraic terms.

Figure~\ref{fig:example:tln:dgs}-(iii) presents a \textit{malformed} \tln{} dynamical graph that is reported invalid by the \tln{} language because it includes a \codein{V-V} connection which introduces unexpected voltage terms into the underlying differential equations. The \tln{} language requires valid \tln{}s to have alternating \codein{I} and \codein{V} nodes to ensure the Telegrapher's equations are faithfully implemented.

\begin{figure*}[!tb]
    \centering
    \subfloat[Branched t-line.]{
        \includegraphics[width=0.24\linewidth]{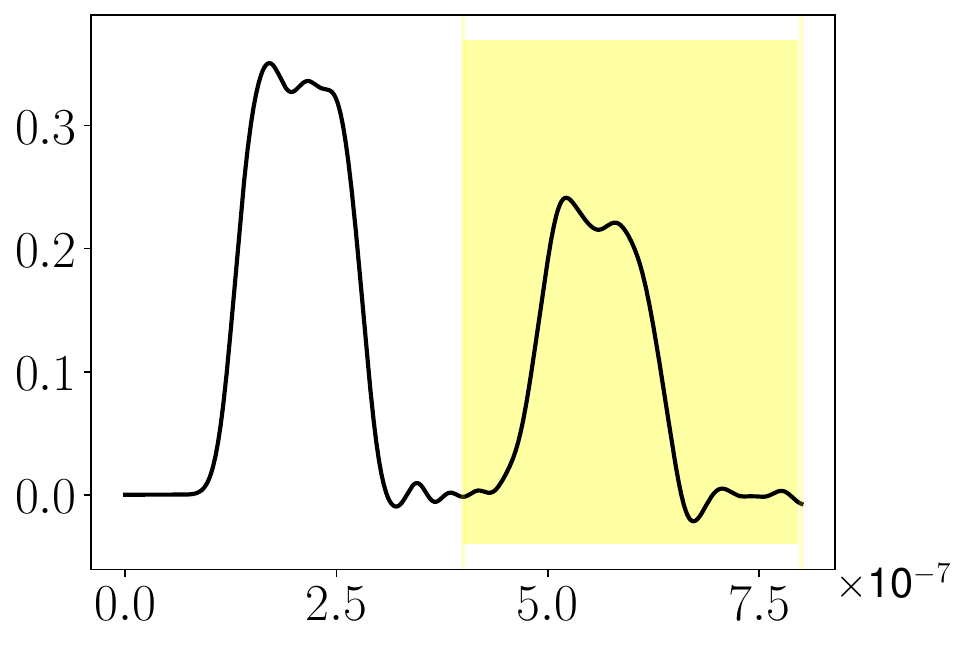}
        \label{fig:example:tln:ideal-branch-traj}
    }
    \subfloat[Linear t-line.]{
        \includegraphics[width=0.24\linewidth]{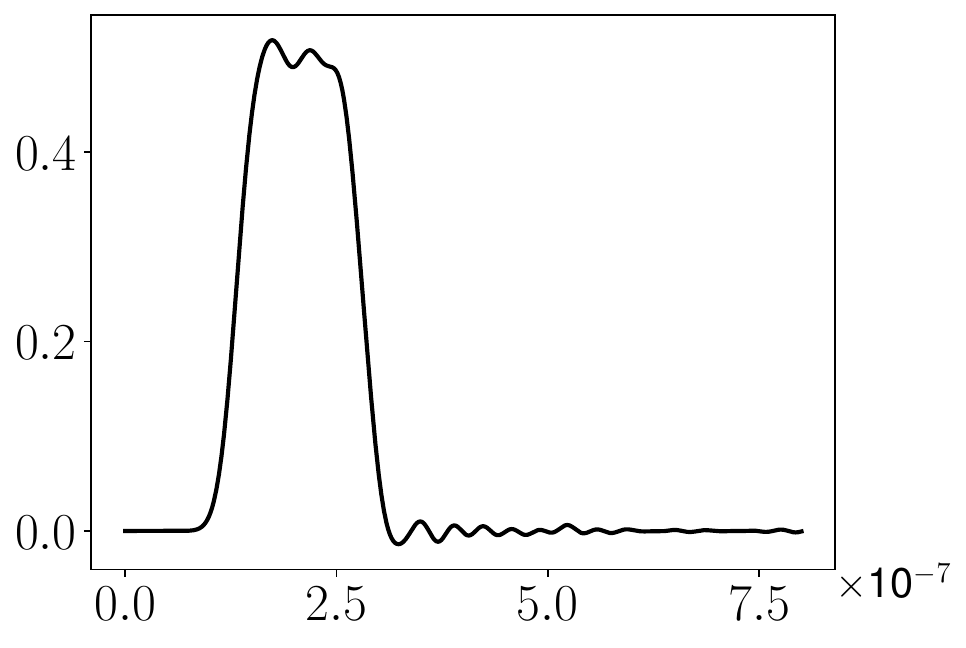}
        \label{fig:example:tln:ideal-linear-traj}
    }
    \subfloat[$\cint$ mismatched t-line.]{
        \centering
        \includegraphics[width=0.24\linewidth]{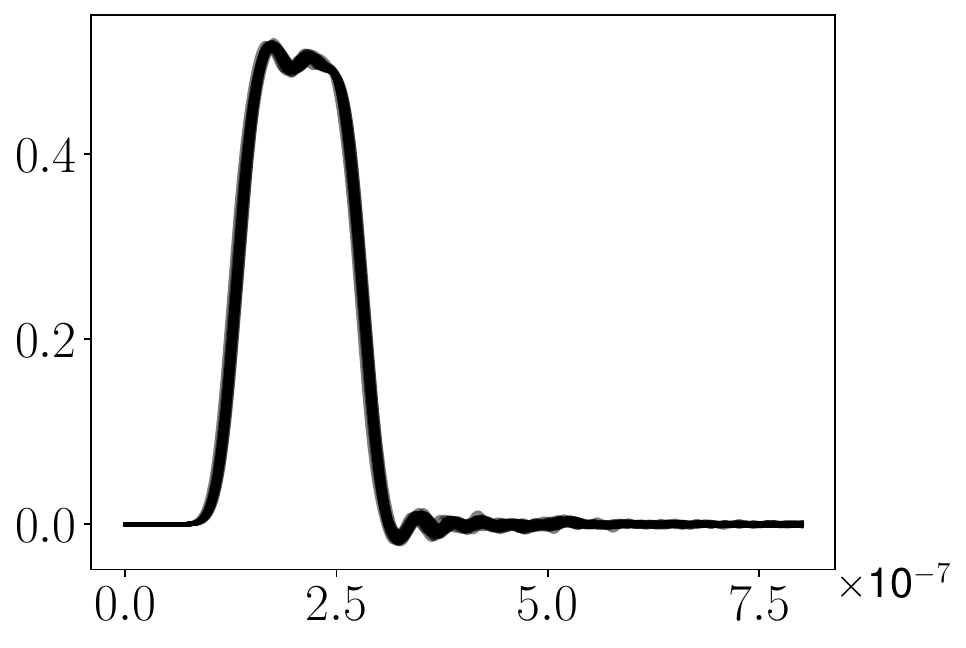}
        \label{fig:example:mm-node-traj}
    }
    \subfloat[$\gm$ mismatched t-line.]{
        \centering
        \includegraphics[width=0.24\linewidth]{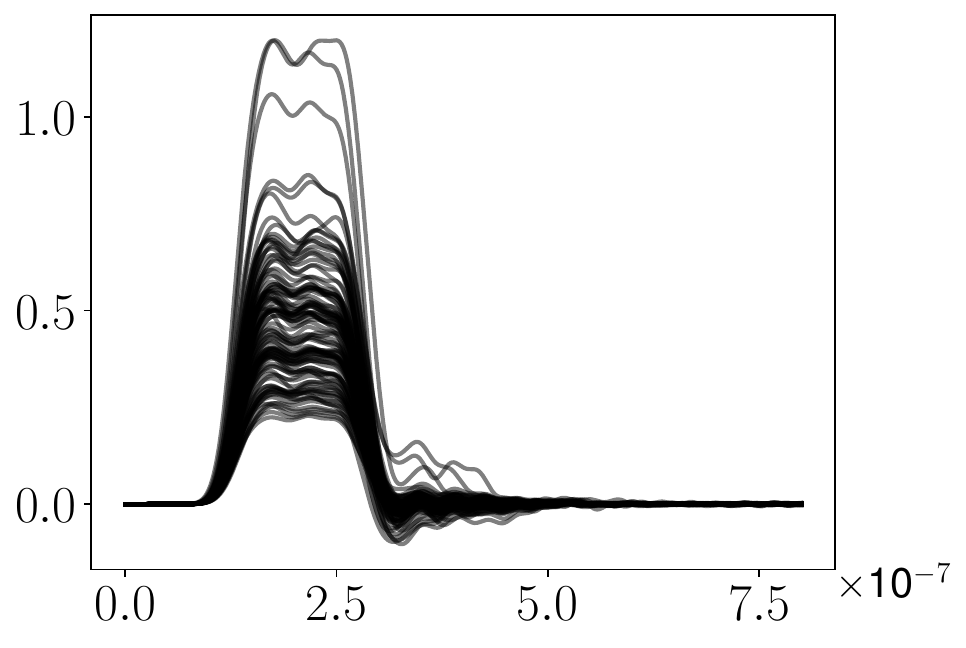}
        \label{fig:example:mm-edge-traj}
    }
    \caption{Dynamics of t-lines observed at \codein{OUT_V}.}
\end{figure*}

\proseheading{Analysis.} We simulate the voltage trajectory\footnote{We simulate 53-node branched and linear lines (Figure~\ref{fig:example:tln:dgs}-(i), (ii)).} at node \codein{OUT_V} for the branched t-line (Figure~\ref{fig:example:tln:ideal-branch-traj}) and a linear, non-branched t-line (Figure \ref{fig:example:tln:ideal-linear-traj}) using differential equations generated by the \tln{} dynamical system compiler. The branched t-line produces a weaker initial pulse (\codein{0.5} $\rightarrow$ \codein{0.3}) and an "echo" of the initial pulse after \codein{4e-8} seconds have elapsed (the shaded area in Figure~~\ref{fig:example:tln:ideal-branch-traj}). This echo occurs because part of the injected pulse travels down the branch and reflects back to the main line, where the pulse then splits and travels to both the \codein{OUT_V} and \codein{IN_V} nodes.  This echoing behavior can potentially be exploited to design PUFs with more complicated system dynamics. 

These trajectories can be used to set signal observation windows. The linear t-line and branched t-line require observation windows of \codein{1e-8} to \codein{3e-8} seconds and \codein{1e-8} to \codein{8e-8} seconds respectively. The branched t-line is assigned a larger observation window to ensure that at least one of the signal echoes is captured in the response encoding. 

\subsection{GmC Circuit Implementation of \tln{} Computing}\label{subsec:example:gmc}

\begin{figure}[!b]
    \centering
    \includegraphics[width=0.95\linewidth]{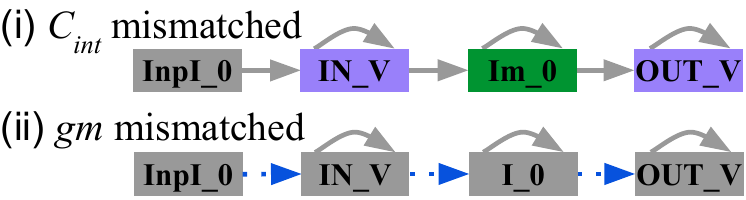}
    \caption{Dynamical graphs of mismatched t-lines.}
    \label{fig:example:mm-dgs}
\end{figure}

A transmission line network is efficiently emulatable with a network of GmC integrators~\cite{khorramabadi1984high}. Each GmC integrator (Figure~\ref{fig:example:gmc-integrator}) consists of transconductors $\gm_1$ and $\gm_2$ (implemented with several transistors) that convert input voltage signals, \(v_{i1}\) and \(v_{i2}\), into currents \(i_{i1}\) and \(i_{i2}\), where \( i_{i1} = \gm_1 \cdot v_{i1}\) and \( i_{i2} = \gm_2 \cdot v_{i2}\). The resistor with conductance \(\gint\) contributes to a current \(i_o\), where \(i_o = \gint \cdot v_o\). The \(i_{i1}\)  \(i_{i2}\), and  \(i_{o}\) currents all flow out of the capacitor \(\cint\), yielding the following dynamics:
\begin{align}
    \frac{dv_o}{dt} = \frac{1}{\cint}(-\gm_1\cdot v_{i1} - \gm_2\cdot v_{i2} - \gint\cdot v_o)
\end{align}
The above dynamics implement the $\frac{dI_i}{dt}$ differential equation from the telegrapher's equation math model, provided \(-\gm_1=\gm_2=\gm\). The \(v_o\), \(v_{i1}\), and \(v_{i2}\) voltages map to $I_i$, $V_{i-1}$ and $V_{i}$ respectively, and the \(\gm/\gint\) and \(\cint/\gm\) circuit quantities implement $G$ and $L$ parameters.
The above dynamics also implement $\frac{dV_i}{dt}$ differential equation; the \(v_o\), \(v_{i1}\), and \(v_{i2}\) voltages maps to $V_i$, $I_{i}$ and $I_{i+1}$ variables and the \(\gint/\gm\) and \(\cint/\gm\)  terms implement $R$ and $C$.

\proseheading{Modified Telegrapher's Equations.} If we relax the \(-\gm_1=\gm_2=\gm\) constraint to allow the device parameter magnitudes \(|gm_1|\), \(|gm_2|\) and \(|gm|\) to differ, then we can introduce \(ws\) and \(wt\) parameters in the \tln{} dynamics and implement a more flexible  version of the telegrapher's equations:
\begin{equation}
    \begin{cases}
         \frac{dV_{i}}{dt} = \frac{1}{C_i}( wt_i \cdot I_{i} - ws_{i+1} \cdot I_{i+1} - G_i\cdot V_{i})\\
        \frac{dI_{i}}{dt} = \frac{1}{L_i} (wt_{i-1} \cdot V_{i-1} - ws_{i} \cdot V_{i} - R_i\cdot I_{i})
    \end{cases}
    \label{eq:example:mod-telegrapher}
\end{equation}
With this relaxed circuit usage, the \(\cint\) and \(\gint\) device parameters implement \(C\)/\(L\) and \(G\)/\(R\) respectively, and the  \(-\gm_1\) and \(\gm_2\) device parameters implement \(wt\) and \(ws\). When \(wt_i=ws_i=1\), the GmC circuit implements \tln{} computing.

\subsection{Exploring Analog Mismatch with \tool{}}\label{subsec:example:mismatch}

The \puf{}'s security properties come, in part, from its sensitivity to fabrication variations (e.g., device mismatch). In a variation-sensitive \puf{} design,  fabricated instances of the same \puf{} behave differently when provided with the same input and can therefore be used to identify an individual uniquely.

We use the \gmctln{} extension to the \tln{} computing model, which codifies the design space of mismatch-sensitive GmC circuit implementations, to study the effects of process variation on a linear t-line. The node types \codein{Vm} and \codein{Im} inherit \codein{V} and \codein{I}, respectively, and override the \codein{C} and \codein{L} attributes to model the random mismatch associated with the corresponding \(\cint\) device parameter. The edge type \codein{Em} inherits \codein{E} and adds two new mismatched attributes \codein{ws} and \codein{wt}, corresponding to the \(\gm\) random mismatch. The \gmctln{} language uses the modified Telegraphers equations to model the dynamics of the \codein{Em} edges. In the \gmctln{}  language, mismatched parameter values are sampled from a normal distribution with a 10\% relative standard deviation before execution.

\proseheading{Analysis.} We use the \gmctln{} language to explore the effects of different sources of process variation on the linear \tln{}. \tool{}'s inheritance system ensures (1) the original linear t-line can be simulated in the \gmctln{} language and deliver the same dynamics, (2) nodes derived from \tln{} language nodes can be substituted into the dynamical graph. The process variation sensitive linear t-lines in Figures \ref{fig:example:mm-dgs} substitute in \codein{Vm}/\codein{Im} node types (pink, green) and \codein{Em} edge types respectively to selectively model the effects of \(\cint\) and \(\gm\) mismatch.

Figures \ref{fig:example:mm-node-traj} and \ref{fig:example:mm-edge-traj} present the \codein{OUT_V} voltage trajectories over 100 sampled device mismatches for the \(\cint\)-sensitive t-line and the \(\gm\)-sensitive t-line respectively. We observe that within the \codein{1e-8} to \codein{3e-8} observation window, the \(\gm\)-sensitive t-line experiences a much greater degree of variation across trials than the \(\cint\)-sensitive t-line. This observation has several implications: (1) future \tln{} architectures should use \(\gm\) mismatch instead of \(\cint\) mismatch; (2) the analog designer should pursue GmC circuit variants that maximize \(\gm\) mismatch -- these variants can be provided to the domain specialist as new edge types. Therefore, domain specialists can use \tool{} to assist in design space exploration and help set the direction of analog design efforts.

\section{The Dynamical Graph Computational Model}\label{sec:dg}

\tool{} employs a higher order computational model, termed a Dynamical Graph (\dg{}) that is specialized to implement different analog computational paradigms. 
The \dg{} contains a set of nodes $\nodeset{}$ and a set of directed edges $\edgeset$ and can be interpreted as a system of differential equations that describes a computation.

\proseheading{Typed Nodes and Edges.}
All elements in the \dg{} are \textit{typed}, meaning each node $\node\in\nodeset$ and edge $\edge\in\edgeset$ belong to a node type $\nodetype$ and an edge type $\edgetype$ respectively.
Node and edge types may define named attributes, and a node type $\nodetype$ has an additional \textit{reduction} operator $\reduction_j \in \set{\sum,\prod}$ and \textit{order} $\nodeorder$ defined.
The type, attributes, order, and reduction are necessary for deriving the graph dynamics.

\proseheading{Dynamics.}
Each node $\node\in\nodeset$  maps to a variable $Q$ in the underlying dynamical system, and each edge $\edge\in\edgeset$ contributes terms to the connected variables' dynamics. 
A node type with an order of $0$ implements a pure function, and a node type with an order of $\nodeorder > 0$ implements $\norder{\nodeorder}$ order differential equations. 
The \dg{} works with a set of production functions $\mathfrak{P_{in}}$, $\mathfrak{P_{out}}$, and $\mathfrak{P_{self}}$ for incoming edges ($\othernode \rightarrow \node$),  outgoing edges ($\othernode \leftarrow \node$), and self-referencing edges ($\circlearrowright \node$) respectively.
Given a node $\node$ of node type $\nodetype$ with $m$ incoming edges, $q$ outgoing edges, and $r$ self-reference edges, the production function takes input as the types of an edge and the source and destination nodes.
The production function finds the production rule that matches the given types and returns an algebraic expression based on the rule and types.
The expressions aggregate over edges with the reduction operator $\reduction$.
Overall, the dynamics of the variable associated with a node are defined as follows:

\begin{small} 
    \begin{align}
        \begin{split}
            \frac{d^{\nodeorder}Q}{dt^{\nodeorder}} =& \ \ \ \  \reduction_{i=0}^{m-1} \mathfrak{P_{in}}(\edgetype_i,\nodetype_i,\nodetype)\ +\reduction_{i=m}^{m+q-1} \mathfrak{P_{out}}(\edgetype_i,\nodetype,\nodetype_i)\ +  \\
            &\reduction_{i=m+q}^{m+q+r-1} \mathfrak{P_{self}}(\edgetype_i,\nodetype)
        \end{split}
    \end{align}
\end{small}

The dynamics of a \dg{} are fully specified with the differential equations collectively from all variables.

\begin{figure}
\footnotesize
\bf
\subfloat{
    \parbox{\linewidth}{
    \centering
        \begin{tabular}{lll}
            \multicolumn{3}{c}{$x \in \realnumbers$, $s \in \positivereal$, $i \in \integers$, $p \in \nonnegativenat$}\\
            \multicolumn{3}{c}{$v \in Literals$, $e \in Expressions$, $b \in BoolExpressions$}\\
            \litl{inherit}($\tuple{Rule}$) & ::= & $Rule$ \syn{inherits} $v$ | $Rule$\\
        \end{tabular}
    }
}\\
\subfloat{
    \parbox{\linewidth}{
        \begin{tabular}{@{\makebox[3em][l]{{\footnotesize\cmt{\texttt{\rownumber}}}\space}}lll}
        $SigT$ & ::= & \syn{real[}$x_0$\syn{,}$x_1$\syn{]} | \hwst{\syn{real[}$x_0$\syn{,}$x_1$\syn{]} \syn{mm(}$s_0$\syn{,}$s_1$\syn{)}} \\
            &&| \syn{int[}$i_0$\syn{,}$i_1$\syn{]} | \syn{lambd}\syn{(}$v$*\syn{)} \\
        $SigTProg$ & ::= & $SigT$ | \hwst{$SigT$ \syn{const}} \\ 
        $Attr$ & ::= & \syn{attr} $v$ \syn{=} $SigTProg$ | \syn{init}\syn{(}$i$\syn{)}  $SigTProg$\\
        $Reduc$ & ::= & \syn{sum} | \syn{mul} \\
        $Type$ & ::= & \syn{node-type(}$p$\syn{,} $Reduc$\syn{)} | \syn{edge-type} \\
        & & | \hwst{\syn{edge-type} \syn{fixed}}\\
        $ProdExpr$ & ::= & $v$ \syn{<=} $e$ | \hwst{$v$ \syn{<=} $e$ \syn{off}}\\
        $ProdRule$ & ::= & \syn{prod}\syn{(}$v_0$ \syn{:} $v_1$\syn{,}$v_2$ \syn{:} $v_3$\syn{->}$v_4$ \syn{:} $v_5$\syn{)}$ProdExpr$\\
        $VAtom$ & ::= & $p$ | \syn{inf}\\
        
        $VMatch$ & ::= & \syn{match}\syn{(}$VAtom$\syn{,}$VAtom$\syn{,}$v_0$\syn{,}$v_n$\syn{->}\syn{[}$v_t$*\syn{]}\syn{)} \\
        &&| \syn{match}\syn{(}$VAtom$\syn{,}$VAtom$\syn{,}$v_0$\syn{,}\syn{[}$v_t$*\syn{]}\syn{->}$v_n$\syn{)}\\
        &&| \syn{match}\syn{(}$VAtom$\syn{,}$VAtom$\syn{,}$v_0$\syn{)}\\
        $ValExpr$ & ::= & \syn{acc} $VMatch$* | \syn{rej} $VMatch$* \\
        $ValRule$ & ::= & \syn{cstr} $v_n$\syn{:}$v_1$ \syn{\{} $ValExpr$* \syn{\}} | \syn{extern-func} $v$\\
        $LangSt$ & ::= & \litl{inherit}($Type$  $v$ \syn{\{}$Attr$*\syn{\}}) | $ProdRule$ | $ValRule$\\
        $LangDef$ & ::= & \litl{inherit}(\syn{lang} $v$ \syn{\{} $LangSt$* \syn{\}}) \\\\
        $Val$ & ::= & $x$ | $i$ | \syn{lambd}\syn{(}$v$*\syn{):} $e$ \\ 
        $FuncVal$ & ::= & $Val$ | $v$\\ 
        $FuncSt$ & ::= & \syn{node} $v_0$ \syn{:} $v_1$ | \syn{edge}\syn{<} $v_0$\syn{,}$v_1$ \syn{>} $v_2$ \syn{:} $v_3$\\
        && | \syn{set-attr} $v_0$\syn{.}$v_1$ \syn{=} $FuncVal$ \\
        && | \syn{set-edge} $v$ \syn{when} $b$ \\
        && | \syn{set-init} $v$\syn{(}$i$\syn{)} \syn{=} $FuncVal$\\
        $FuncArg$ & ::= & $v$ \syn{:} $SigT$ | $v_0$\syn{.}$v_1$ \syn{:} $SigT$\\\\
        $FuncDef$ & ::= & \syn{func} $v_0$\syn{(}$FuncArg$*\syn{)} \syn{uses} $v_1$ \syn{\{}$FuncSt$* \syn{\}}\\\\
        $Stmt$ & ::=& $FuncDef$ | $LangDef$\\
        $Prog$ & ::= & $Stmt$*
        \end{tabular}
    }
}
\caption{Basic grammar for \tool{} language. Shaded expressions model hardware behavior.}
\label{fig:grammar}
\end{figure}

\section{The \tool{} Programming Language}\label{sec:cdg:overview:speclan}

Figure~\ref{fig:grammar} presents the \tool{} programming language. \tool{} languages specialize the \dg{} computational model to implement different classes of analog computations, \tool{} functions generate dynamical graphs and can be invoked to generate \dg{}s with different topologies and attribute parametrizations. Sections~\ref{sec:lang:langdef}-\ref{sec:lang:func} describes the basic \tool{} language and function definition constructs, and \ref{sec:lang:analog} describes the \tool{} hardware extensions.
Section~\ref{subsec:lang:tln-example}-\ref{subsec:lang:gmc-tln} provides an illustrative example with \tln{}.

\noindent\textbf{Expressions.} Expressions are defined over variables, which include nodes, edges, and function arguments (\codein{$v$}), simulation time (\codein{\syn{time}}), and node and edge attributes \codein{$v$\syn{.}$v'$} (attribute $v'$ of node $v$). Boolean expressions  $b$ use logical and comparison operators and return boolean values. Math expressions ($e$) use linear/non-linear math operators (e.g., $+, \times, sin, cos$, etc) and if-then-else statements (e.g. \codein{\syn{if} b \syn{then} $e$ \syn{else} $e'$}) and return real values. \tool{} checks that all variables in the variable are in scope and that expressions evaluate to their expected type.

\noindent\textbf{Datatypes.} \tool{} supports bounded real and integer datatypes (\codein{\syn{real}[$x_0$,$x_1$]} and \codein{\syn{int}[$i_0$,$i_1$]}), and function datatypes \codein{\syn{lambd}($v$*)} that accept real-valued arguments and compute real values.  \tool{} checks variable assignments to ensure the value matches the variable's datatype, and is contained within the datatype's value range \codein{[$x_0$,$x_1$]}. All assigned functions  \codein{\syn{lambd}($v$*): $e$} must have the same number of arguments as the function datatype,

\subsection{Language Definitions (Lines 1-17)}\label{sec:lang:langdef}

Each language definition declares node and edge types and defines production rules and validation rules over node and edge types. The typing information and production rules are used to specialize the dynamical graph computational model to implement the desired analog computational paradigm.

 \proseheading{Types.} \lines{[Lines 1-7]} Each \codein{\syn{node-type} $v$($p$,$Reduc$) $Attr$*} statement defines a node type  \codein{$v$} with a variable order \codein{$p$} and a reduction operator \codein{$Reduc$}. The node type \codein{$v$} contains attributes and initial value definitions $Attr$* that specify the names and datatypes of attributes and initial values. Each \codein{\syn{edge-type} $v$ $Attr$*} statement defines an edge type named \codein{v} with attributes $Attr$*. Each attribute and initial value may optionally be assigned a constant value.

\proseheading{Production Rules.} \lines{[Lines 8-9]} Each \codein{\syn{prod}} statement defines a new production rule for a connection with an edge $v_0$ of  type $v_1$, a source node $v_1$ of type $v_3$, and a destination node $v_4$ of type $v_5$. The production expression (\codein{$v$ <= $e$}) applies a term to the node $v$. A production rule is a self-referencing rule if the source and destination node have the same name ($v_2$=$v_4$).

\proseheading{Local Validity Rules.} \lines{[Lines 10-14]} \tool{} supports the definition of local validity rules over the cardinalities of connected node and edge types (\codein{\syn{cstr}}). Each local validity rule \codein{\syn{cstr} $v_n$:$v_1$ {ValExpr*}} operates on a target node named $v_n$ of node type $v_1$ and returns true if the accepted validity expression (\codein{\syn{acc}  $VMatch*$}) returns true and the rejected validity expression (\codein{\syn{rej} $VMatch*$}) returns false. Both expressions return true if the target node $v_n$ is \textit{described} by the pattern $VMatch*$.
Here, we say that a node is described by a pattern if we find an assignment of edges to clauses $VMatch$ in the pattern such that every clause is assigned between $VAtom_1$ and $VAtom_2$ edges of edge type $v_0$ connected to a node type contained in \codein{[$v_t$*]}.
\tool{} provides rejected validity expression to support restricting the design space as an accepted expression can become invalid once a language extension is introduced.

\proseheading{Global Validity Rules.} \lines{[Line 14]} \tool{} supports the definition of a global validity check that is evaluated over the entire graph topology. The \codein{\syn{extern-func} $v$} statement binds an external validity checking algorithm $v$. Global connectivity checks are  required to ensure the \dg{} implements certain topologies, such as grid topologies.

\proseheading{Semantic Checks.} \tool{} ensures all node and edge types have unique names,  that each node type contains an initial value declaration \codein{\syn{init}($i$) $SigTProg$} for derivatives $0...i..p-1$ of the node type, and ensures edge types contain only attribute statements. For each production rule,  \tool{} checks the expression $e$ only references variables instantiated in the \codein{\syn{prod}(.)} clause and that the node $v$ in the production expression is either source node $v_2$ or the destination node $v_4$ from the clause. For each validity rule, \tool{} ensures match clauses reference the target node $v_n$ and that all node and edge types are declared in the language.

\subsubsection{Inheritance}\label{subsec:lang:inheritance}

\tool{} supports single inheritance of languages, where the derived language inherits node and edge types, production rules, and validation rules from a parent language and may define node and edge types that inherit from types in the parent language. \tool{} constrains the derived language to ensure the parent and derived languages are compatible:

\begin{itemize}

\item Derived node and edge types inherit the parent type's node order and reduction operator, and inherit all attributes and initial value declarations from the parent type.

\item Inherited attributes and initial values can be redefined in the derived node or edge type but must retain the same datatype (real, integer, lambda) and operate on a smaller value range than the parent attribute.

\item Production and validation rules from the parent class cannot be overridden or removed. Any new production or validation rules must include one new type from the derived class.

\item For each connection, the most specific production rule is applied for a given combination of derived nodes and edge types. If no production rule exists, \tool{} falls back to a production rule that applies to the parent node and edge types. Ambiguities also produce an error.
\end{itemize}

The above properties ensure dynamic graphs comprised of derived types can be cast to the parent type, and dynamical graphs written in the parent language can be faithfully executed in the derived language and progressively rewritten to use node and edge types in the derived language.

\subsection{Function Declarations (Lines 19-27)}\label{sec:lang:func}

\tool{} supports defining functions that procedurally generate dynamic graphs from a set of function inputs. Each function declaration  \codein{\syn{func} $v_0$ ($FuncArg$*) \syn{uses} $v_1$} specifies the \tool{} language $v_1$ to use and a list of typed arguments $FuncArg$* accepted by the function. 

The function body contains statements that construct a dynamical graph parametrized over the function arguments. The \codein{\syn{node} $v_0$ : $v_1$} statement constructs a node named $v_0$ of type $v_1$, and the \codein{\syn{edge}<$v_0$,$v_1$> $v_2$ : $v_3$} statement constructs an edge with name $v_2$ of type $v_3$ that connects the source node $v_0$ to the destination node $v_1$. The \codein{\syn{set-attr} $v0$.$v1$ = $FuncVal$} statement sets an attribute $v_1$ of node $v_0$ to a value or function argument, and the \codein{\syn{set-init} $v$($i$) = $FuncVal$} statement sets the initial value of the \norder{i} derivative of node $v$ to value or function argument. The \codein{\syn{set-switch} $v$ \syn{when} $b$} statement turns the switchable edge $v$ on when $b$ is true, where $b$ is a boolean expression over function arguments.

\proseheading{Semantic Checks} \tool{} checks that all referenced node and edge types are defined in the language, all referenced nodes/edges are defined within the function body, and all attributes and initial values defined in the node/edge type are set for each node, and all datatype assignments are valid.

\subsection{\tool{} Hardware Extensions}\label{sec:lang:analog}

\tool{} offers language extensions (grey highlighted in Figure~\ref{fig:grammar}) for modeling analog-specific behaviors not already captured in the basic language features. The \codein{\syn{real}[$x_0$,$x_1$] \syn{mm}($s_0$,$s_1$)} datatype models process variation-sensitive attributes and initial values, and samples a mismatched value $\hat x$ from a normal distribution $N(x,x \cdot s_0 + s_1)$ when an attribute or initial value is assigned to a nominal value $x$. Each function invocation sets the random seed used to produce the same mismatched values. The seed can be varied across invocations to model multiple fabricated instances of a particular design.

\tool{} offers language constructs for defining non-programmable attributes and initial values (\codein{\syn{attr} $v$...\syn{const}} and \codein{\syn{init-val}($i$)...\syn{const}}) and fixed edge types (\codein{\syn{edge-type} \syn{fixed} $v_1$}). Non-programmable switches are always on, and non-programmable attributes must be assigned to a constant value or math function on instantiation. \tool{} also supports defining production rules (\codein{\syn{prod}... $v$ <= $e$ \syn{off}}) that model nonidealities associated with edges that are switched off.

\proseheading{Semantic Checks}  For attributes and initial values that are assigned to function arguments, \tool{} checks that the associated definition in the type declaration is not \codein{\syn{const}}. \tool{} validates that all \codein{\syn{set-switch}} statements are applied to edges that are not \codein{\syn{fixed}}.

\begin{figure}[t]
\input{gen-specs/lang-tln-language.tex}
\caption{\tool{} \tln{} language definition snippet. The \codein{ntyp} and \codein{etyp} are abbreviations of \codein{node-type} and \codein{edge-type} respectively}
\label{example:tln:langdef}
\end{figure}

\subsection{Illustrative Example: \tln{} Language}\label{subsec:lang:tln-example}

We present an example \tln{} language definition in~Figure~\ref{example:tln:langdef}. The \tln{} language declares \codein{V} and \codein{I} node types, which map to $V$ and $I$ terms in the discretized Telegrapher's equations, and \codein{InpV} and \codein{InpI} node types, which provide external input voltage and currents into the \tln{} computation. The \codein{V} and \codein{I} node types contain real-valued \codein{c}/\codein{g} and \codein{l}/\codein{r} attributes which map to the $C$, $G$, $L$, and $R$ parameters in the Telegrapher's equations. 

The \tln{} language defines production rules that derive the Telegrapher's equations from node and edge types. For example, the \codein{\syn{prod}(e:E, s:V->t:I)} production rule matches an edge type \codein{E} that connects a source node \codein{s} of type \codein{V} to a destination node \codein{t} of type \codein{I}, and contributes \codein{$-var(t)/s.c$} term to the source node \codein{s}'s dynamics\footnote{\codein{var(.)} is a convenient function that returns the state variable associated with the node.}. The \codein{V} node type implements a first-order differential equation with a summing reduction operator, so the \codein{$-var(t)/s.c$} term is added to the derivative of \codein{$s$}.

The \tln{} language definition includes validation rules that ensure the \tln{} computational model is faithfully implemented. The validation rule \codein{\syn{cstr} I} disallows connections between \codein{I} and \codein{I} node types and the validation rule \codein{\syn{cstr} V} disallows connections between \codein{V} and \codein{V} node types.


\begin{figure}
\centering
\input{gen-specs/lang-tln-br-func}
\caption{\tool{} function snippet of branched T-Line}
\label{example:tln:func}
\end{figure}

\proseheading{Branched T-Line Function.} Figure~\ref{example:tln:func} presents an \tool{} function that implements a programmable version of the branched t-line from Figure~\ref{fig:example:tln:dgs}-(i). The \codein{br-func} function accepts a \codein{br} branch bit and enables the edge connecting the \codein{IN_V} and \codein{I_2} nodes together if the bit is set. Invoking the function \codein{br-func} with \codein{br=0} and \codein{} returns the dynamical graph of a linear t-line (Figure~\ref{fig:example:tln:dgs}-(ii)) and a branched t-line respectively. The body of the \codein{br-func} function uses the \tln{} language to construct the branched t-line topology (\codein{\syn{uses} tln}). In the above function, all resistances and conductances are set to zero, all capacitances and inductances are set to \codein{1e-09}, and the \codein{InpI_0} is configured to provide a trapezoidal pulse function with width \codein{2e-8} at time \codein{t=0}.

\subsection{The \gmctln{} Language}\label{subsec:lang:gmc-tln}

Figure~\ref{example:tln:gmctln} presents the \gmctln{} language, which models the nonidealities of a mismatch-sensitive GmC network. The \gmctln{} language inherits all types and rules from the \tln{} language. All \tln{} computations are implementable in the \gmctln{} language and deliver the same dynamics.

The \gmctln{} language defines \codein{Vm} and \codein{Im} node types that inherit from the \codein{V} and \codein{I} node types and incorporate the effects of device mismatch on the \(\cint\) parameter in the corresponding GmC circuit. The \codein{Vm} and \codein{Im} node types override the \codein{c} and \codein{l} attributes to accept values between \codein{1e-10} and \codein{1e-08} and are subject to 10\% mismatch. When a value $v$ is written to an \codein{Im} node's mismatched attribute \codein{c},  \codein{c} is set to a value sampled from a normal distribution $N(v, 0.1v)$. Because the \codein{Vm} and \codein{Im} node types inherit from the \codein{V} and \codein{O} node types, they can be used anywhere a \codein{V} node type was originally used.

The \gmctln{} language also defines a mismatched edge \codein{Em} that both incorporates the effects of device mismatch on the \(gm_1\) and \(gm_2\) device parameters in the \codein{GmC} circuit. The \codein{Em} edge inherits from the \codein{E} edge and defines 10\% mismatched \codein{ws} and \codein{wt} attributes that map to the \(gm_1\) and \(gm_2\) device parameters. Because the GmC circuit dynamics change when \(gm_1 \neq gm_2\), the \gmctln{} language defines new production rules for the \(Em\) edge. These production rules implement the modified Telegrapher's equations presented in Section~\ref{subsec:example:gmc}.

\proseheading{Empirical Validation.}  We randomly generate 1000 valid \gmctln{} \model{}s and generate SPICE netlists from these models with a simple algorithm. We observe (1) all valid \model{}s successfully map to a spice-level netlist,  (2) the generated transient dynamics of the \model{} match the transient dynamics of the spice-level netlist within a root-mean-squared error of 1\%. Therefore, empirically, the \gmctln{} language captures the dynamics of the spice-level circuit.

\begin{figure}[t!]
\centering
\input{gen-specs/lang-gmc-tln-language}
\caption{Gm-C-\tln{} language definition snippet.}
\label{example:tln:gmctln}
\end{figure}

\subsection{The \tool{} Framework}

Given an \tool{} program containing language and function definitions, an end user may invoke any of the defined functions with \tool{}. \tool{} executes the function with the provided arguments to build the associated dynamic graph and then validates that the dynamic graph satisfies the local and global validation rules in the associated language (Section~\ref{sec:dg:validator}). If the dynamic graph validates, \tool{} generates differential equations (Section~\ref{sec:dg:compiler}) that simulate the transient behavior of the graph.

\section{\tool{} Dynamical System Compiler}\label{sec:dg:compiler}
\begin{algorithm}[!tbp]
\small
\caption{Differential Equation Compilation}\label{alg:dg:compilation}
    \begin{algorithmic}[1]
    \Require $dg, langDef$
    \Ensure A system of equations describes the $dg$ dynamics
    \State $eqs \gets []$
    \State $eq.\algapp(\algkw{InitState}(dg))$
    \For{$n$ \algkw{in} $\algkw{Nodes}(dg)$}
        \State $eqs.\algapp(\algkw{LowOrdEqs}(langDef, n))$
        \State $rhs \gets []$
        \For{$e$ \algkw{in} $\algkw{Edges}(n)$}
            \State $rule \gets \algmatchprod(lagnDef, n, e)$
            \State $expr \gets \algkw{Rewrite}(rule, n, e)$
            \State $rhs.\algapp(expr)$
        \EndFor
        \State $eq \gets \algkw{FormEq}(n, rhs)$
        \State $eqs.\algapp(eq)$
    \EndFor
    \State \Return $eqs$
    \end{algorithmic}
\end{algorithm}
The \tool{} compiler processes a dynamical graph and a language definition as input and generate differential equations. 
For each node with order \(p\), the compiler generate \(p\) differential equations with each state variable \(n_i, i=1,...p\) corresponding to the $\norder{i}$ derivative of the node \(n\). 
The compilation process, outlined in Algorithm~\ref{alg:dg:compilation}, proceeds as follows:
First, the compiler creates all the necessary state variables with initial values specified in the dynamical graph.
The \algfnprose{LowOrdEqs} function generates equations \(\ddt{n_i} = n_{i+1}, i=1,...p-1\) which describe the node dynamics for orders less than \(p\).

Subsequently, the compiler iterates through each edge associated with the nodes to determine the $\norder{p}$ derivative of the node. 
The \algfnprose{LookUpProdRule} function examines the node and edge to find an associate production rule (\codein{\syn{prod}($v_0$:$v_1$, $v_2$:$v_3$->$v_4$:$v_5$) $ProdExpr$}) in the language definition.
Specifically, the edge's type determines $v_1$, and the edge's direction and the terminal nodes' types determine $v_3$ and $v_5$, uniquely identifying the production rule to apply.
If no rule matches immediately, the compiler will trace the inheritance relation and look up parent types recursively to find the closest production rule to apply.
The \algfnprose{Rewrite} function retrieves the expression ($ProdExpr$) from the rule and returns new expression with the nodes and edges in the $ProdExpr$ substituted by the ones currently processed.
The expressions of all edges of the node $n$ are aggregated in \algfnprose{FormEq} using the reduction operator specified in the node type, yielding the equation representing the $\norder{p}$ derivative of the node.
Following the procedure, the compiler returns differential equations \(eqs\), which fully specify the computation of the given dynamical graph and can be used for transient simulation.

\section{\tool{} Dynamical Graph Validator}\label{sec:dg:validator}

The \tool{} validator takes input as a dynamical graph and a language definition, performing the validation required outlined in Section~\ref{sec:lang:langdef}.
The dynamical graph is provided as an input to the external function (\codein{\syn{extern-func} $v$}) specified in the language definition, which validates that the graph satisfies all global validity rules. The graph validates if the function \codein{v} returns true.
The validator verifies the local validity rules by iterating through all nodes and checks if they are described by at least one accepted pattern (\codein{\syn{acc} $VMatch$*}) and not described by any of the rejected patterns (\codein{\syn{rej} $VMatch$*}).

We formulate and solve the \textit{described} relation as an Integer Linear Programming (ILP) problem in Algorithm~~\ref{alg:dg:validation-ilp}.
The algorithm accepts as input a node \(n\) and a pattern which is essentially a list of clauses (\codein{$VMatch$*}), and returns whether the pattern describes the node.
The procedure initializes an empty list of constraints \(cstrs\) and ILP variables \(vars\) 
The variables denote the assignment of edges to clauses, meaning if \(vars[i][j]=1\), the $\norder{i}$ edge of the node \(n\) is assigned to the $\norder{j}$ clause of the pattern.
Every edge-clause pair is inspected using the \algfnprose{Matched} function, which returns true if the edge is of edge type \codein{$v_0$} connected to a node type contained in \codein{[$v_t$*]} specified in \codein{$VMatch$}.
If true, \(var[i][j]\) can be assigned to the clause, meaning it is constrained to be either 0 and 1; \(var[i][j]\) is constrained to equal 0 otherwise.
The constraints specified in \algfnprose{UnityRowSum} ensure that each edge is assigned to only one clause, i.e., \(\forall i.\sum_{j}var[i][j] = 1\).
The \algfnprose{RangeColSum} encodes the cardinality constraints specified with the \codein{VAtom} terms hold for each clause, i.e., \(\forall j. VAtom_{j,1} <= \sum_{i}var[i][j] <= VAtom_{j,2}\).
The constraints are passed to an ILP solver which returns true if a satisfying assignment is found and returns false otherwise.

\begin{algorithm}[t!]
\small
\caption{Pattern Matching}\label{alg:dg:validation-ilp}
    \begin{algorithmic}[1]
    \Function{IsDescribed}{$n, pattern$}
        \State $cstrs\gets[]$
        \State $vars\gets\algkw{ILPVars}(\algkw{len}(edges)\times\algkw{len}(pattern))$
        \State $edges\gets\algedges(n)$
        \For{$i, e$ \algkw{in} $\algkw{enumerate}(edges)$}
            \For{$j, cls$ \algkw{in} $\algkw{enumerate}(pattern)$}
                \If {$\algkw{Matched}(n, e, cls)$}
                    \State $cstrs.\algapp(\algkw{ZeroOrOne}(vars[i][j]))$
                \Else
                    \State $cstrs.\algapp(\algkw{Zero}(vars[i][j]))$
                \EndIf
            \EndFor
        \EndFor
        \State $cstrs.\algapp(\algkw{UnityRowSum}(vars))$
        \State $cstrs.\algapp(\algkw{RangedColSum}(pattern, vars))$
        \State \Return $\algkw{ILPSolve}(cstrs)$
    \EndFunction
    \end{algorithmic}
\end{algorithm}

\section{Evaluation}
We present two case studies where we formalize analog compute paradigms and the tradeoffs associated with different analog realizations with \tool{}. In each case, we analyze the effect of incorporating a subset of analog behaviors into the computation and formulate topological constraints that guide different analog design problems.
\tool{}'s goal is to enable iterative co-design flow of domain-specific unconventional compute paradigms and analog circuits.
Therefore, expressiveness and extensibility are the overarching objectives of the system.

\subsection{Cellular Nonlinear Network (\cnn)}
\begin{figure}[!b]
    \centering
    \subfloat[][\codein{sat} (\textcolor{blue}{blue}) and \codein{sat_ni} (\textcolor{orange}{orange}).]{
        \centering
        \includegraphics[width=0.5\linewidth]{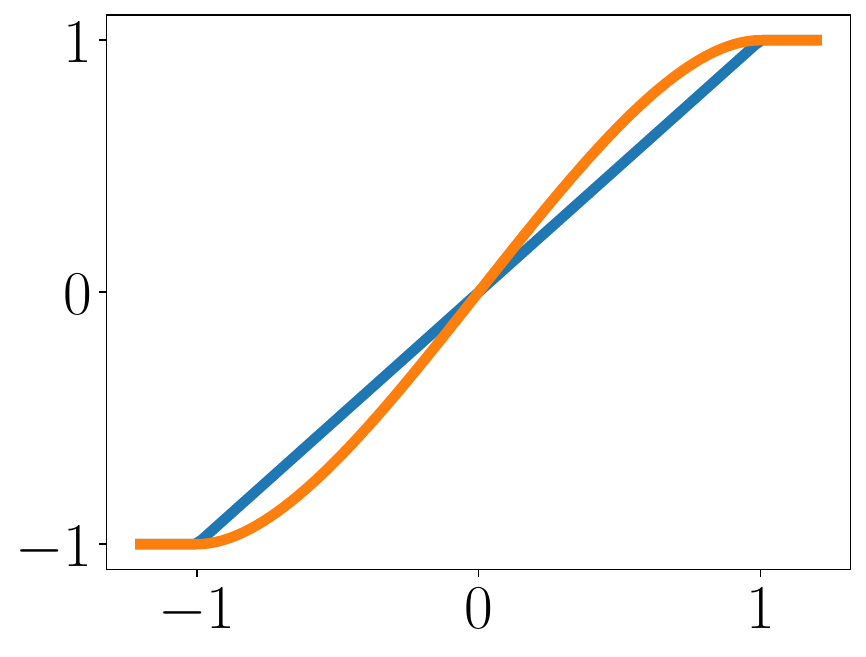}
        \label{fig:eval-saturation-cmp}
    }
    \subfloat[][Input image.]{
        \centering
        \includegraphics[width=0.35\linewidth]{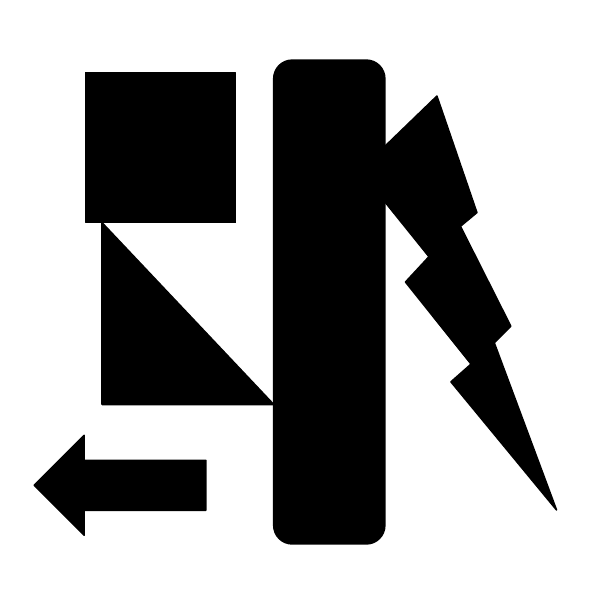}
        \label{fig:eval-cnn-input}
    }\\
    \centering
    \subfloat[width=\linewidth][Simulation results of an edge detection \cnn{} with different types in \codein{cnn} and \codein{hw-cnn} language modeling hardware non-idealities. 
    \cnnexp{A}: Ideal CNN. \cnnexp{B}: 10\% random mismatch in integrator bias. \cnnexp{C}: 10\% random mismatch in parameters. \cnnexp{D}: Non-ideal saturation function]{
        \parbox{\linewidth}{
            \centering
            \includegraphics[width=0.7\linewidth]{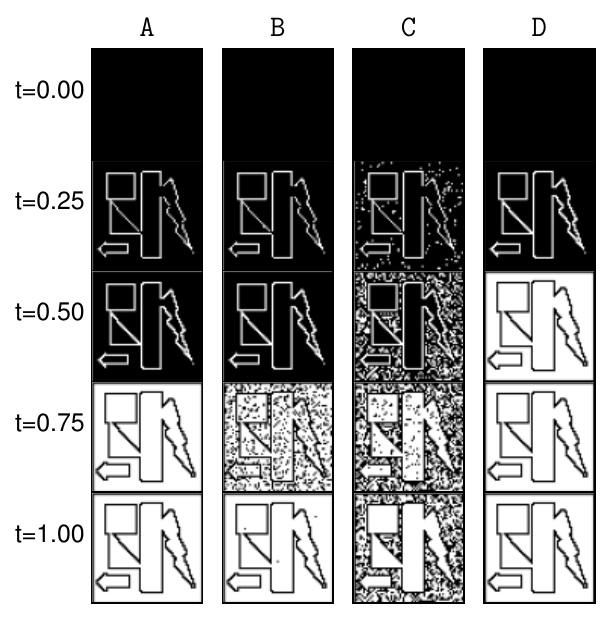}
        }
        \label{fig:eval-cnn-output}
    }
    \caption{Saturation functions and input image to \cnn{} in the experiments and the observed transient dynamics under different non-ideal conditions.}
\end{figure}

The cellular nonlinear network~\cite{chua1988cellular-theory} analog compute paradigm performs computation over a topology of locally interconnected cells, and has applications in image processing, pattern recognition, PDE solving, and security primitives~\cite{chua1988cellular-theory, chua1995cellular-pde, fortuna2001cellular-processing, csaba2009cnn-puf}. Researchers have previously developed analog accelerators that implement the \cnn{} computational paradigm~\cite{cruz199816cnn-chip, duan2014memristor}. The following differential equation describes the dynamics of a \cnn{} with cells $x_{ij}$:
\begin{equation}\label{eq:eval:cnn}
    \ddt{x_{ij}} = -x_{ij} + \sum\limits_{(k, l)\in N(i, j)}(A_{ij,kl}\cdot f(x_{kl}) + B_{ij,kl}\cdot u_{kl}) + z 
\end{equation}
 Each cell \(x_{ij}\) and accepts an external input \(u_{ij}\). The  neighboring cells ($k,l \in N(i,j)$) and associated external inputs flow into cell $x_{ij}$. The $A_{ij,kl}$ and $B_{ij,kl}$ matrices apply weights to the neighboring signals and external inputs, and the nonlinear activation function $f$ transforms neighboring signals. Each $x_{i,j}$ also has negative feedback and is subject to a constant bias $z$. The cell's non-linear activation function $f$ is typically a saturating function (blue line, Figure~\ref{fig:eval-saturation-cmp}).

\begin{figure}[!t]
\subfloat[\cnn{} language.]{
    \parbox{\linewidth}{
        \input{gen-specs/lang-cnn-language}
        \label{fig:eval:cnn-lan}
    }
}\\
\subfloat[HW-\cnn{} language.]{
    \parbox{\linewidth}{
        \input{gen-specs/lang-hw-cnn-language}
        \label{fig:eval:hw-cnn-lan}
    }
}
\caption{\tool{} \cnn{} language definition and extension.}
\label{fig:eval:all-cnn-lan}
\end{figure}

\proseheading{The \cnn{} Language.} Figure~\ref{fig:eval:cnn-lan} presents the \tool{} \cnn{} DSL. The \codein{V} and \codein{Inp} node types map to the $x_{i,j}$ and $u_{i,j}$ variables, and the \codein{Out} node type applies the nonlinearity dynamic \codein{act} to an incoming $x_{i,j}$ signal. The \codein{iE} and \codein{fE} edge types implement the \cnn{} dynamics. The \codein{V} node type defines the \codein{z} parameter, and the \codein{fE} edge type defines the \codein{g} attribute which implements the \(A\) and \(B\) parameters.

\proseheading{Hardware Extensions.} Figure~\ref{fig:eval:hw-cnn-lan} presents the \codein{hw-cnn} extension to the \cnn{} language that codifies the analog \cnn{} design space and models circuit nonidealities.\cite{fernandez2009cnn-mismatch,Razavi2005analog-design} The \codein{Vm}, and \codein{fEm} node types extend \codein{V} and \codein{fE} respectively and override the \codein{g} and \codein{z} attributes to incorporate mismatch in the hardware realization -- this nonideality is reported to affect system 
convergence in analog implementations~\cite{fernandez2009cnn-mismatch}. The \codein{OutNL} node type inherits from the \codein{Out} node type and applies a non-ideal saturation function \codein{sat\_ni} with nonlinear dynamics near the saturation points (orange line, Figure~\ref{fig:eval-saturation-cmp}). These non-idealities arise because analog \cnn{} realizations implement saturation with a MOS differential pair which introduces non-linearities due to the MOS transistors' large signal behavior~\cite{Razavi2005analog-design}.

\proseheading{Edge detection.} We implement an edge detector~\cite{chua1988cellular-app} in the \codein{cnn} language and then use the \codein{hw-cnn} language extension to explore the effect of different analog non-idealities. The edge detector \cnn{} is provided input image pixels as an external $u_{i,j}$ input. The value of each cell $x_{i,j}$ at steady state computes the output pixel $p_{i,j}$, which is black if an edge is detected. Figure~\ref{fig:eval-cnn-input} presents the input image and column A of Figure~\ref{fig:eval-cnn-output} presents the expected output image for the edge detector. Columns B-D present the \cnn{} edge detector's behavior with integrator bias (\codein{z} mismatch), \codein{g} parameter mismatch, and non-ideal saturation behavior. These behaviors are modeled by selectively substituting \codein{V}, \codein{Out}, and \codein{fE} nodes in the original implementation with non-ideal nodes from \codein{hw-cnn}. The rows of Figure~\ref{fig:eval-cnn-output} capture the evolution of time.

\proseheading{Analysis.} We observe all analog nonidealities substantially affect the transient dynamics of the edge detector. 
Notably, designs with mismatched \codein{z} and \codein{g} parameters (B, C) converged more slowly, where \codein{g} mismatch also yielded an incorrect output image. The non-ideal saturation function (D) produced the correct result and actually \textit{improved} convergence time, suggesting that some nonidealities have a potentially positive effect on the computation model. 

From this analysis, we can conclude (1) analog \cnn{} designs should prioritize reducing \codein{g} mismatch over \codein{z} mismatch, (2) introducing certain nonidealities into the non-linearity function is acceptable, and in some case beneficial. The first point informs designers of where to expend effort to improve fidelity, and the second point suggests a potential axis to explore in the \cnn{} design space. Overall, this analysis underscores the importance of early-stage modeling of nonidealities in the design process, and how studying these nonidealities can aid in design-space exploration.

\begin{figure}[!t]
\subfloat[\con{} language.]{
    \parbox{\linewidth}{
        \input{gen-specs/lang-obc-language}
        \label{fig:eval:obc-lan}
    }
}\\
\subfloat[Ofs-\con{} language.]{
    \parbox{\linewidth}{
        \input{gen-specs/lang-ofs-obc-language}
        \label{fig:eval:ofs-obc-lan}
    }
}
\caption{\tool{} \con{} language definition and extension.}
\label{fig:eval:all-obc-lan}
\end{figure}

\subsection{Oscillator-Based Computing (\con)}

Oscillator-based computing (\con) is an emerging unconventional compute paradigm that has recently been used to solve \maxcut~\cite{chou2019Con, ochs2021Con, mallick2021con-global} and graph coloring problems~\cite{mallick2021con-global} and perform signal processing tasks like filtering and pattern recognition~\cite{vodenicarevic2017con,csaba2020coupled}. In oscillator-based computing, a network of coupled oscillators performs computation. The coupling strength encodes the program inputs, and the synchronization behavior between oscillators implements the computation. The phase dynamics of the oscillators are described by the \textit{modified Kuramoto model}~\cite{ochs2021Con}:

\begin{equation}
    \ddt{\phi_i} = - C_1\cdot\sum\limits_{j=1}^{n}K_{ij}\cdot\sin{(\phi_i-\phi_j)} -C_2\cdot\sin{(2\phi_i)} 
\end{equation}
The \(\phi_i\) variable is the phase of oscillator $i$, and the \(K_{ij}\cdot\sin{(\phi_i-\phi_j)}\) term models the coupling between oscillators \(i\) and \(j\), with \(K_{ij}\) denoting the coupling strength. $C_1$ and $C_2$ are constant scaling factors. We use \codein{1.6e9} and \codein{1e9} respectively in the evaluation.

\proseheading{The \con{} Language} Figure~\ref{fig:eval:obc-lan} defines an \con{} language with \tool{}. The \codein{Osc} node type and \codein{Cpl} edge type models the oscillator phase \(\phi\) and the oscillator coupling behavior respectively. The \codein{Cpl} defines an attribute \codein{k} that maps to the coupling strength \(K\). The production rules implement the modified Kuramoto model where the constants $C_i$ and $C_j$ are embedded in the expression.

\proseheading{Integrator Bias Hardware Extension} Figure~\ref{fig:eval:ofs-obc-lan} presents the ofs-\con{} hardware extension to the \con{} language, which models the effect of analog integrator offset on an integrator-based \con{} accelerator design. Prior work uses integrator and nonlinear conductors to implement the phase dynamics~\cite{ochs2021Con}, which is possible to experience a non-zero offset when fed by currents emulated the coupling function. The \codein{Cpl_ofs} node type inherits the \codein{Cpl} node type and defines a mismatched offset attribute \codein{offset} that biases the coupling terms. The ofs-con language defines new production rules that capture the effect of this offset nonideality on the system dynamics.

\begin{table}[t!]
    \centering
    \footnotesize{}
    \begin{tabular}{l|cr|cr|}
        \cline{2-5}
         & \multicolumn{2}{c|}{\codein{obc}} & \multicolumn{2}{c|}{\codein{offset-obc}} \\ \hline
        \multicolumn{1}{|c|}{\(d\)} & \multicolumn{1}{c|}{\begin{tabular}[c]{@{}c@{}}sync \\ prob. (\%)\end{tabular}} & \multicolumn{1}{c|}{\begin{tabular}[c]{@{}c@{}}slvd\\ prob (\%)\end{tabular}} & \multicolumn{1}{c|}{\begin{tabular}[c]{@{}c@{}}sync\\ prob. (\%)\end{tabular}} & \multicolumn{1}{c|}{\begin{tabular}[c]{@{}c@{}}slvd\\ prob. (\%)\end{tabular}} \\ \hline
        \multicolumn{1}{|l|}{\(0.01\pi\)} & \multicolumn{1}{r|}{94.1} & 94.1 & \multicolumn{1}{r|}{54.1} & 54.1 \\ \hline
        \multicolumn{1}{|l|}{\(0.1\pi\)} & \multicolumn{1}{r|}{94.2} & 94.1 & \multicolumn{1}{r|}{94.8} & 94.6 \\ \hline
    \end{tabular}
    \caption{Probability of successful synchronization and solving \maxcut{} problems with \con{} and \con{} with offset.}
    \label{tab:eval:con-cmp}
\end{table}

\proseheading{Max-Cut Solver.} We implement a \maxcut{} solver as a function using the \con{} language, and then substitute \codein{Cpl} with  \codein{Cpl_ofs} nodes to study the effect of integrator bias on the max-cut solution. We invoke the maxcut solver function over 1000 unweighted 4-vertex graphs, where the \maxcut{} solver maps input edges to coupling strengths and graph nodes to oscillators. For each graph, the oscillator phases are extracted from the \maxcut{} solver simulation at steady-state. Oscillator nodes with phases that are within \(d\) radians of \(0\) and \(\pi\) are placed in partition 0 and partition 1, respectively; oscillators with other phases are marked unknown. The \(d\) deviation tolerance parameter is configurable and external to the analog circuit.

Table~\ref{tab:eval:con-cmp} summarizes the simulation results. 
For the \codein{obc} \maxcut{} solver, the correct partition is returned 94\% of the time. Conversely, the \maxcut{} solver with the integrator bias nonideality only successfully partitions graphs 54.1\% percent of the time \(d\). We study the dynamics of the nonideal \maxcut{} solver and find the oscillator phase experiences slight jitter. We increase the phase tolerance from \(0.01\pi\) to \(0.1\pi\) to absorb this jitter and find the non-ideal \maxcut{} solver attains 94\% accuracy with this new parametrization. In this study, we used \tool{} to analyze analog non-idealities, which were then attenuated by applying a compensation technique external to the circuit. This mitigation approach allowed us to significantly improve the \maxcut{} solver without actually improving the fidelity of the underlying circuit.

\begin{figure}[t]
\input{gen-specs/lang-intercon-obc-language}
\caption{Intercon-\con{} language definition.}
\label{fig:eval:intercon-obc-lan}
\end{figure}

\begin{figure}[t]
\input{gen-specs/lang-mm-tln-language}
\caption{Intercon-\con{} language definition.}
\label{fig:eval:intercon-obc-lan}
\end{figure}

\proseheading{Modeling Routing Tradeoffs.} Figure~\ref{fig:eval:intercon-obc-lan} presents the intercon-\con{} extension to the \con{} language, which captures the design  tradeoffs associated with different kinds of coupled oscillator interconnect -- a critical aspect in \con{} design. The language lets end-users intermix \textit{all-to-all} connections~\cite{mallick2021con-global} and neighboring connections in the \con{} computations.~\cite{ahmed2021con-neighbor} All-to-all connections offer exceptional programmability but require significantly larger area as more programmable interconnect is required. Neighboring connections are less flexible but significantly more resource-efficient.  The impact of this particular design decision is significant: the all-to-all chip~\cite{mallick2021con-global} implemented 30 oscillators and devoted most area to routing circuitry, while the neighboring connection chip~\cite{ahmed2021con-neighbor} implemented 560 oscillators with minimal routing circuitry. Both chips were fabricated with the CMOS 65nm technology and used 1.44 mm\textsuperscript{2} chip size.

The intercon-\con{} language formalizes the trade-off between programmability and resource utilization. This language defines two edge types \codein{Cpl\_l} and \codein{Cpl\_g}, which implement edges for local and global connections, respectively. The \codein{Cpl\_l} and \codein{Cpl\_g} edges define a \codein{cost} attribute, which specifies the resource cost of a connection. Local edges are assigned a lower cost than global edges. The \con{} language defines validation rules that mandate connections across groups to be realized using \codein{Cpl\_g} edges. These rules ensure the connectivity restrictions associated with local and global edge types are enforced at compile-time. With intercon-\con{}, we can soundly architect global-local interconnect topologies that capture programmability/efficiency trade-off points in the design space.

\section{Related Works}

\proseheading{Empirical Studies}: Researchers have characterized the impact of non-idealities on solution quality, convergence time, and stability for cellular nonlinear networks~\cite{fernandez2009cnn-mismatch} and oscillator-based computing~\cite{vodenicarevic2017con, chou2019Con, ahmed2021con-neighbor, ochs2021Con}.
These studies involve experts with a deep understanding of both the computing paradigm and analog circuitry and present the paper as a technical artifact. Our work complements this research and focuses on codifying both analog compute paradigms and the analog circuit constraints with a unified representation accessible to both parties, enabling collaboration between domain specialists and analog designers.


\noindent\textbf{Analog Models.} Analog behavioral modeling languages such as Verilog-A and Verilog-AMS~\cite{pecheux2005vhdl, liao2014verilog, wang2008behavioral, rhim2015verilog} and  Analog MacroModels~\cite{wang2007parameterized,de2005mixed, ajayi2020open} are typically used to model analog behavior for system-level design. These models are usually constructed from a transistor-level circuit design and apply simplifications or elide certain analog behaviors (e.g., transient behavior) to improve the performance of the model and focus on the modeling of classical circuits for circuit designer's use. In contrast, our method targets early design-space exploration for the co-design of novel compute paradigms and analog circuits, which require accessible models for non-circuit domain specialists and support of iterative modeling over circuit specification before a transistor-level circuit is presented.

\noindent\textbf{Structured Math Abstractions.} Many fields use structured math representations of continuous-time systems, such as signal-flow graphs and block diagrams~\cite{holtz1995representation, ki1998signal, veerachary2004general}. These representations  compose basic math elements (e.g., multiplication, filtering) together to construct a higher-level computation. These approaches are hardware agnostic and do not support easy incorporation of analog design constraints and nonidealities. In contrast, \tool{} captures both analog design tradeoffs and restrictions and models the overall dynamics of the continuous-time system.

\noindent\textbf{Analog Computer Specifications.} Compiler-writers have also developed analog hardware specification languages to capture the capabilities of GPAC analog computing platforms~\cite{achour2016arco, achour2018jaunt, achour2020Legno}. These languages cannot effectively model analog compute paradigms with interacting elements (e.g., \con{}, \cnn{} models), and rely on algorithms to implement certain hardware constraints (e.g., current fanout). Our language can support a range of compute models, including compute models with interacting elements, and supports the specification of the aforementioned interconnect restrictions.

\section{Conclusion}
Reconfigurable analog circuits are promising substrates for unconventional compute paradigms tailored to specific domains. 
We present the \tool{} programming language for specifying these novel compute paradigms while incorporating analog constraints and trade-off space.
\tool{} enables progressive modeling of analog behaviors with computations and provides a validator and dynamical system compiler for verifying and simulating the computations.
We describe three unconventional computing paradigms with \tool{} and demonstrate that \tool{} aids design space exploration and co-design of the computation with the analog circuits.
Therefore, \tool{} takes a key step towards an agile design of novel compute paradigms, enhancing collaboration among experts from diverse backgrounds, and thereby stimulating innovation.

\section*{Acknowledgement}
We would like to thank Boris Murmann and Luke Sammarone for their input and expertise during development of the \tool{} language.
This work is supported by the U.S. Department of Energy, Office of Science, Office of Advanced Scientific Computing Research under Award Number DE-FOA-0002950 and the Stanford Graduate Fellowship.

\bibliographystyle{plain}
\bibliography{references}

\end{document}